\documentclass[aps,pra,
amsmath,amssymb,
reprint
]{revtex4-1}

\usepackage[left=1.5cm, right=1.5cm, top=1.785cm, bottom=2.0cm]{geometry}
\usepackage{mathptmx}

\usepackage{amsthm}
\usepackage[utf8]{inputenc}
\usepackage{xspace}
\usepackage{units}
\usepackage[disable]{todonotes}

\newcommand{\Ecut}{E_\text{cut}}
\newcommand{\Ecutt}{E_\text{cut}^{(2)}}

\newcommand{\opnorm}[1]{\ensuremath\left\|#1\right\|_\text{op}}
\newcommand{\norm}[1]{\ensuremath\left\|#1\right\|}

\newcommand{\dftk}{DFTK\xspace}
\newcommand{\julia}{Julia\xspace}

\newcommand{\nl}{{\rm nl}}
\theoremstyle{definition}
\newtheorem{theorem}{Theorem}

\newcommand\exactWF{{u}}
\newcommand\trialWF{{\widetilde u}}
\newcommand\exactEV{{\varepsilon}}
\newcommand\trialEV{{\widetilde\varepsilon}}
\newcommand\res{{\widetilde r}}

\newcommand\bk{{\mathbf k}}
\newcommand\br{{\mathbf r}}
\newcommand\bb{{\mathbf b}}
\newcommand\bG{{\mathbf G}}
\newcommand\bq{{\mathbf q}}
\newcommand\cR{{\mathcal R}}
\newcommand\R{{\mathbb R}}
\newcommand\Z{{\mathbb Z}}
\newcommand\N{{\mathbb N}}
\DeclareMathOperator*{\argmin}{argmin}

\begin{document}
\title{
	A posteriori error estimation for the
	non-self-consistent Kohn-Sham equations%
}

\author{Michael F. Herbst}
\email{michael.herbst@inria.fr}
\affiliation{
	CERMICS, Ecole des Ponts and Inria Paris, 6 \& 8 avenue Blaise Pascal, 77455 Marne-la-Vall\'ee, France
}

\author{Antoine Levitt}
\email{antoine.levitt@inria.fr}
\affiliation{
	CERMICS, Ecole des Ponts and Inria Paris, 6 \& 8 avenue Blaise Pascal, 77455 Marne-la-Vall\'ee, France
}

\author{Eric Cancès}
\email{eric.cances@enpc.fr}
\affiliation{
	CERMICS, Ecole des Ponts and Inria Paris, 6 \& 8 avenue Blaise Pascal, 77455 Marne-la-Vall\'ee, France
}

\begin{abstract}
  We address the problem of bounding rigorously the errors in the numerical solution of
  the Kohn-Sham equations due to (i) the finiteness of the basis set, (ii) the
  convergence thresholds in iterative procedures, (iii) the propagation of
  rounding errors in floating-point arithmetic. In this contribution, we
  compute fully-guaranteed bounds on the solution of the non-self-consistent
  equations in the pseudopotential approximation in a plane-wave basis set. We
  demonstrate our methodology by providing band structure diagrams of silicon
  annotated with error bars indicating the combined error.
\end{abstract}

\maketitle
\section{Introduction}

Experimental results are provided with error bars in most scientific
fields. In an ideal world, this should also be the case for results
obtained by numerical simulation. Complementing simulation results
with error bars is becoming mandatory in some branches of engineering,
such as aeronautics or car industry, in which simulation has
partially, or even totally, replaced experiment (e.g.~virtual wind
tunnels~\cite{windtunnel} or crash simulators~\cite{carcrash}).
Obviously, uncontrolled errors in the numerical simulations used
to design and test an aircraft are likely to have dramatic
consequences for passengers and crews.

What about molecular simulation?
{\it Statistical} error bars are often displayed in molecular dynamics (MD) and
quantum Monte Carlo (QMC), where stochastic processes (e.g.~Langevin equation
for MD, drift-diffusion stochastic differential equations for QMC) are at the
core of the simulation. 
However, these error bars are purely statistical in nature and are not
guaranteed: they only reflect the finiteness of the statistical sample and the
actual solution of the model has a positive probability to lay outside the
confidence interval. In addition, they only take into account one of the
components of the error between the exact value of the quantity of interest
(q.o.i.) and its
numerical approximation. The other components of the error are deterministic in
nature and are also present in simulations of purely deterministic models, such
as those based on density-functional theory (DFT) which are dealt with in the article.

Consider as an example of q.o.i.~the lattice constant of Silicon at
\unit[0]{K}.
In order to compute a numerical approximation of this q.o.i., we
first have to select a model. We have at our disposal an outstanding
model: the many-body Schr\"odinger equation with relativistic
corrections. However, solving this equation directly is completely out of
reach. The first approximation is to replace this reference model with
a cruder, but tractable, reduced model.
We consider as a reduced model the one obtained by successively using
the Born-Oppenheimer approximation to decouple as much as possible
nuclear and electronic degrees of freedom, and the Local Density
Approximation of the Kohn-Sham density-functional theory (KS-LDA),
together with a pseudopotential model to avoid representing core
electrons and the singularities of the orbitals near atoms. This gives
rise to a well-defined mathematical model, hopefully with a unique
solution. We refer to the difference between this approximation and
the physical reality as the {\em model error}. 

The reduced model, though simpler than the reference model, still has an
infinite number of degrees of freedom, and must be discretized to be simulated.
For the crystalline phase, a typical
(simplified) workflow is as follows~\cite{martin2004electronic}:
\begin{enumerate}
\item The infinite computational domain is truncated to a finite supercell with periodic boundary conditions,
  numerically handled through Brillouin zone sampling;
\item In this supercell, crystalline orbitals are discretized using a
  finite basis set of
  plane waves;
\item The self-consistent Kohn-Sham equations are solved by a
  self-consistent field (SCF) algorithm;
\item At each step of the SCF algorithm, a linear eigenvalue
  equation is solved with an iterative eigensolver;
\item All computations are performed with finite precision.
\end{enumerate}
We refer to the error in steps 1 and 2 as the discretization error, to
the error in steps 3 and 4 as the algorithmic error, and to the error
in step 5 as the arithmetic error. To these must be added programming
errors,
not to be neglected in codebases consisting of millions of
lines of code, and hardware errors,
which are expected to become
significant for exascale architectures~\cite{snir2014addressing}.
These two latter kinds of errors we will not treat.

Note that all steps mentioned above are systematically improvable:
by increasing parameters,
such as Brillouin zone sampling, basis set cutoff, convergence
thresholds, or using higher-precision arithmetic,
results get more accurate at the price of longer computation times.
The usual procedure for controlling these errors is to perform convergence
studies:
on the system of interest or a related system with similar
characteristics, the parameter is increased until the variation of the
quantity of interest is below a specified tolerance. Over time,
knowledge of ``good'' parameter values solidifies into rules of thumb
that are automated in codes or suggested to users in manuals. When
used appropriately, this results in acceptable errors that are
below the target accuracy (for instance, for pseudopotential methods,
the error compared to
all-electron models)~\cite{Lejaeghere2014,lejaeghere2016reproducibility}.

This empirical process is however still problematic. First, it might result
in suboptimal performance by excess of caution. Second, it still
requires a degree of hand-tuning and is thus problematic for fully
automated computations, which are useful for \textit{in silico} design
of novel compounds and for building databases of material or chemical
properties. Third, the rules of thumb can fail for unusual systems with
unexpected behavior, i.e.~exactly those where accurate simulations
are important to aid understanding.

The purpose of \textit{a posteriori} error analysis is to automate and
rationalize this process by providing accurate, computable and
guaranteed bounds on the error of each step. The purpose of this is
twofold. First, bounds of the total error on the q.o.i.~can be
obtained by simply summing up the different components of the error,
allowing one to bound the accuracy of the final result. Second, the
computer resources necessary to reach a given accuracy on the final
result can be optimized by {\em error balancing} techniques. For
instance, convergence thresholds should be adapted to the
discretization: if the discretization is coarse, it is not necessary
to use extremely tight convergence criteria, since the final accuracy will be
limited by the discretization error anyway. For the same reason, it
might suffice to perform most operations in single-precision
arithmetic to save computational time without losing much on the
accuracy of the final result. Error balancing based on {\it a
  posteriori} error bounds allows one to turn these common sense
remarks into black-box numerical strategies. The user provides the
target accuracy and the software automatically chooses, in an adaptive
way along the iterative process, the reduced models, discretization
bases, convergence thresholds, and data structures to obtain the
desired accuracy at a quasi-optimal computational cost.

Applying this methodology to electronic structure calculations is a challenge
due to the complexity of the equations. Despite this, significant progress has
been made in the past decade towards rigorous error control, which makes this
perspective realistic in the medium term. Let us mention in particular recent
works on {\it a priori} and {\it a posteriori} discretization error bounds for DFT%
~\cite{Cances2012,Zhou2013,Chen2015,chen20152,Lin15,JCP,CDMSVclusters},
including $k$-point sampling~\cite{cances2018numerical}, and on the numerical
analysis of SCF
algorithms~\cite{chupin2020convergence,Zhuang,Rohwedder2011,ZBJ}.

In this contribution, we provide for the first time a combined
analysis of the errors in step 2 (basis set truncation), 4 (inexact
solution of the eigenvalue problem) and 5 (arithmetic error). The
q.o.i.~is the band diagram of silicon, or more precisely the energies
of the 3 highest occupied and 4 lowest unoccupied crystalline orbitals
for specific $k$-points.
The models under consideration are the
Cohen-Bergstresser model~\cite{Cohen-Bergstresser} on the one-hand,
and the non-self-consistent (electron-electron interaction completely neglected)
periodic KS-LDA model with GTH pseudopotentials~\cite{GTH,HGH} on the
other hand,
for which we will give fully guaranteed error bounds.

Some aspects of our analysis, such as the computation of the residual,
are general and can be extended to other quantities of interest, basis
sets and models; others, such as the gap estimation, rely more
specifically on properties of the setup considered. We refer to the
conclusion for more details on this point. Our analysis is, however,
limited in its present state due to the neglect of the errors in steps
1 (Brillouin zone sampling) and 3 (self-consistency). We hope to
address both these challenging aspects in future publications.

Let us finally point out that the model error arising from the choice of
density-functional theory is not easily systematically improvable.
For wavefunctions methods (e.g.~coupled-cluster methods~\cite{Rohwedder2013})
guaranteed {\it a posteriori} error bounds on the model error
--- with respect to the many-body Schr\"odinger equation (MBSE)
--- can in principle be derived,
since a residual of the MBSE can be computed
from the approximate wavefunction and energy.
For DFT, on the other hand, this is not easily the case such that
guaranteed model errors are probably out of reach and
going beyond machine-learned confidence intervals obtained from big data
sets of reference calculations seems very difficult.
Regarding pseudopotentials
and related approaches, the Projector Augmented Wave~\cite{PAW} (PAW) method
should be amenable to the derivation of guaranteed error bounds - with respect
to reference full-electron DFT calculations - since approximate all-electron
Kohn-Sham orbitals can be reconstructed from PAW pseudo-orbitals allowing the
construction of a residual.

\section{Kohn-Sham density-functional theory in the pseudopotential approximation}

As already mentioned in the introduction, the q.o.i.~under
consideration are the energies of crystalline orbitals at specific
$\bk$-points, in a non-self-consistent pseudopotential model. In the
following we denote by $\mathcal R$ the periodic lattice,
and assume that the Bloch wavevector $\bk$ is fixed in the
Brillouin zone. The equation we solve is
\begin{align}
  \label{eq:eigenvalue_eq}
  H \exactWF = \exactEV \exactWF, \quad \int_\Omega |\exactWF(\br)|^2 \, d\br=1,
\end{align}
with $u$ an $\mathcal R$-periodic function and $\Omega$ the unit cell. The periodic
Hamiltonian is given by
\begin{align*}
  H = \frac 1 2 (-i\nabla + \bk)^{2} + V,
\end{align*}
where the (possibly nonlocal) effective potential $V$ is
assumed to be known.%

The problem \eqref{eq:eigenvalue_eq} is discretized in a Fourier basis:
any ${\mathcal R}$-periodic function can be expanded on the orthonormal plane-wave
basis
\begin{align*}
  e_{\bG}(\br) = \frac{1} {\sqrt{|\Omega|}} e^{i \bG\cdot \br}
\end{align*}
where $|\Omega|$ is the volume of the unit cell, and $\bG$ belongs to the
reciprocal lattice $\mathcal R^{*}$. The complete set
$(e_{\bG})_{\bG \in \mathcal R^{*}}$ is truncated to obtain a finite
approximation space 
\begin{align*}
  X = {\rm Span}\left\{e_{\bG}, \frac 1 2 |\bG+\bk|^{2} \le \Ecut\right\}
\end{align*}
with dimension $N_b:=\mbox{dim}(X)$, for some finite cutoff energy $\Ecut > 0$. We will use the convenient abuse of notation consisting in writing $\bG \in X$ to denote a $\bG \in \mathcal R^{*}$ such that $e_{\bG} \in X$. The linear eigenvalue
equation \eqref{eq:eigenvalue_eq} is then discretized by invoking the
variational principle: the $N_b \times N_b$ complex Hermitian matrix
\begin{align*}
  \langle  e_{\bG} | H | e_{\bG'}\rangle = \frac 1 2 |\bG+\bk|^{2} \delta_{\bG\bG'} + \langle  e_{\bG} | V | e_{\bG'}\rangle, \quad \bG,\bG' \in X,
\end{align*}
is diagonalized using an iterative eigensolver~\cite{saad,kresse1996efficient},
employing fast Fourier transform to perform
matrix-vector products efficiently.

The potential $V$ can be a purely local potential, or have a nonlocal
component in the case of norm-conserving pseudopotentials. We will
consider two cases here, chosen for their simple analytic forms.

First we consider the Cohen-Bergstresser
pseudopotentials for semiconductors in the diamond and zinc-blende
structure~\cite{Cohen-Bergstresser}. These are extremely purely local
$\mathcal R$-periodic potentials with a small number of non-zero Fourier
coefficients. More precisely
\begin{align}
  \langle e_{\bG} |V | e_{\bG'}\rangle  = \frac 1{|\Omega|}  \widehat v_{\rm CB}(\bG-\bG') \qquad \mbox{(Cohen-Bergstresser)}, \label{eq:PPCB}
\end{align}
where $\widehat v_{\rm CB}(\Delta \bG)$ is only nonzero for
$\Delta \bG$ in a small finite set (the first five shells of reciprocal
vectors). The coefficients $\widehat v_{\rm CB}(\Delta \bG)$ were adjusted to reproduce
spectroscopic data.

We next consider more realistic Goedecker--Teter--Hutter (GTH) norm-conserving
potentials~\cite{GTH,HGH}.
These potentials are composed of a local and a non-local part:
\begin{align*}
  V = V_{\rm loc} + V_{\nl} \qquad \mbox{(Goedecker--Teter--Hutter, GTH)},
\end{align*}
where
\begin{align}
  \langle e_{\bG} |V_{\rm loc} | e_{\bG'}\rangle &= \frac 1{|\Omega|}  \widehat v_{\rm loc}(\bG-\bG'), \nonumber \\
  \langle e_{\bG}| V_{\nl} | e_{\bG'}\rangle &= \sum_{a}\sum_{lm} \sum_{ij} d_{almij} p_{almi}(\bk+\bG) \overline{p_{almj}(\bk+\bG')}.
  \label{eq:Vnl}
\end{align}
Here $a$ runs over all the atoms in the unit cell, the range of
angular momentum $l$
depends on the chemical element considered, $m = -l, \dots, l$
and the projection indices $i, j$ are summed over a small number of integers
(two in the case of silicon).
The coefficients $\widehat v_{\rm loc}(\Delta \bG)$ are finite sums of products of three terms: a structure
phase factor depending on the position of the atoms in the unit cell, a radial
Gaussian envelope and a radial rational function.
The $p_{almi}({\mathbf q})$
are the product of a structure phase factor, a spherical harmonics
$Y_{lm}$, a radial Gaussian envelope and a radial polynomial of degree $l$.
Among the many types of pseudopotentials available, we chose this type
because their analytic form (rather than tabulated data) lends itself well to rigorous error analysis.

All computations presented in this paper have been performed using the
density-functional toolkit~(\dftk)~\cite{DFTKweb}, a recent
Julia~\cite{Julia} implementation for Kohn-Sham DFT and related
methods using plane-wave basis sets. Our implementation of the
discussed estimates and the code producing the figures of this paper
is available on Github~\cite{reproducers}.

\section{From residuals to errors}
Assume for simplicity that $\exactEV$ is an isolated eigenvalue of
$H$. Given a finite $\Ecut$, the result of the iterative procedure is
a normalized vector $\trialWF \in {\mathcal H}$ and a Riesz approximation
$\trialEV = \langle \trialWF |H| \trialWF \rangle$ of the
eigenvalue $\trialEV$, such that the algebraic residual
$P_X(H\trialWF - \trialEV \trialWF)$, where $P_X$ is the orthogonal projector on $X$, is small. Note that the
algebraic residual vanishes for an exact matrix eigensolver, in
contrast with the global residual
$\res=H \trialWF - \trialEV \trialWF$, which is always nonzero due to
the finite basis discretization. The question we are interested in is:
can we bound rigorously the error between $\trialEV$ and $\exactEV$?

Several {\it a posteriori} error bounds for elliptic eigenvalue
problems are available in the literature%
~\cite{Geor_He_eigs_89,Bab_Osb_eigs_91,Heu_Ran_a_post_FE_eig_02,Mehr_Mied_adpt_eigs_11,Kuzn_Rep_eigs_13,Bank2013-zj,Cars_Ged_LB_eigs_14,Liu_fram_eigs_15,Canc_Dus_Mad_Stam_Voh_eigs_nonconf_18},
the most accurate of them requiring lower bounds on the distance between the
eigenvalue or cluster of
eigenvalues\cite{Gallistl2014-nq,Gallistl2015-yf,Dai2015-jq,Bonito2016-sv,Boffi2017-fj,CDMSVclusters}
of interest and the rest of the spectrum. We focus here on two basic
bounds for the sake of simplicity: the implementation of more
accurate ones is work in progress.
\begin{theorem}[Bauer-Fike]
  There exists an eigenvalue $\exactEV$ of $H$ such that
  \begin{align*}
    |\trialEV-\exactEV| \le \|\res\|.
  \end{align*}
\end{theorem}

This bound is very easy to handle: it only requires an upper bound on
the ${\mathcal H}$-norm of the residual. It is however not sharp.
In particular, it is well-known that
the error on isolated eigenvalues of Hermitian matrices behaves
as the square of the residual.
An {\it a posteriori} error bound having this property is the following.
\begin{theorem}[Kato-Temple]
  Let $\exactEV$ be the eigenvalue of $H$ closest to $\trialEV$. Assume that $\exactEV$ is an isolated point of the spectrum of $H$
  with a distance $\delta > 0$ to the rest of the spectrum. Then
  \begin{align}  \label{eq:KT}
     |\trialEV-\exactEV|  \le \frac{\|\res\|^{2}}{\delta}.
  \end{align}
\end{theorem}
The proof of both these theorems can be found in standard textbooks~\cite{saad}.

To apply these bounds, we need three ingredients:
(i) construct an upper bound on the residual $\norm{\res}$,
(ii) construct a lower bound on the gap $\delta$,
(iii) implement these bounds in the presence of roundoff errors.
We address these problems in sequence.

\section{Computing the residual}
\label{sec:residual}
Assuming that the variational numerical method provides an approximate normalized eigenfunction 
$$
\trialWF = \sum_{\bG \in X} \widetilde u(\bG) e_\bG \in X,
$$
with
$$
\widetilde u(\bG):=\langle e_{\bG}|\widetilde u\rangle \quad \mbox{and} \quad \|\trialWF\|^{2}= \sum_{\bG \in X} |\widetilde u(\bG)|^2=1,
$$
and a Riesz  approximation
$\trialEV=\langle\trialWF|H|\trialWF\rangle$ of the associated
eigenvalue, then the square norm of the residual
$$
\res=H \trialWF - \trialEV \trialWF
$$
can be decomposed as
\begin{equation}
\begin{aligned}
  \|\res\|^2&= \|P_{X} \res \|^{2} + \|P_{X^{\perp}} \res \|^{2}\\
  &= \|P_X(H\trialWF- \trialEV \trialWF)\|^{2} + \|P_{X^{\perp}} V \trialWF\|^{2}
\end{aligned}
	\label{eqn:Residuals}
      \end{equation}
where $P_{X}$ and $P_{X^{\perp}}$ are the orthogonal projectors
on $X$ and $X^{\perp}$ respectively. Here we have used that the
kinetic energy operator is diagonal in reciprocal space, so that
$P_{X^{\perp}} H \widetilde u = P_{X^{\perp}} V \widetilde u$.
The first term, which is easily computed in the Fourier basis of $X$,
is the square of the norm of the in-space, algebraic residual
$P_X(H\trialWF- \trialEV \trialWF)$. It is driven to zero by the
iterative eigensolver but does not vanish because the iterations stop
when the convergence thresholds are reached. This is the origin of the
algorithmic error, and is easily computed explicitly. The second term
is the out-of-space residual and is the source of the discretization
error. This term is more difficult to compute, and most often in
practice only an upper bound can be obtained at a reasonable
computational cost.

\subsection{Cohen-Bergstresser model}
\begin{figure}[h]
\centering
  \includegraphics[width=\columnwidth]{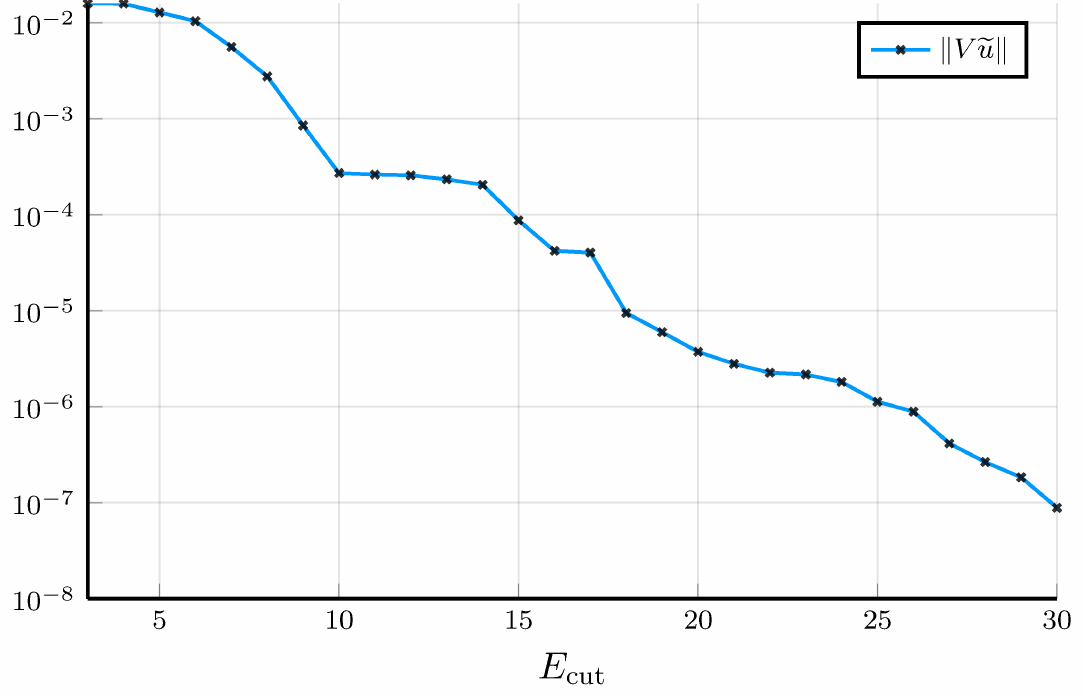}
  \caption{
	  Exact residual norm
	  for the first eigenvalue at the $\Gamma$-point
	  of the Cohen-Bergstresser model of silicon.
	}
  \label{fig:cb_residual}
\end{figure}

In this model, $V$ is given by~\eqref{eq:PPCB}, and only a small
number of terms $\widetilde v_{\rm CB}(\Delta \bG)$ are non-zero.
In this case
\begin{align*}
\|P_{X^{\perp}} V \trialWF\|^{2} = \frac 1 {|\Omega|^{2}} \sum_{\bG \in X^{\perp}}\left|  \sum_{\bG' \in X} {\widehat v_{\rm CB}}(\bG-\bG') \trialWF(\bG') \right|^{2}.
\end{align*}
This computation
extends over a finite range of $\bG$. Denoting by $G_{\rm max}$ the
norm of the largest non-zero Fourier mode of $\widehat v_{\rm CB}$, these can
be identified to be of the form $\bG+\Delta \bG$ for
$\frac 1 2 |\bk+\bG|^{2} \le \Ecut$ and
$|\Delta \bG| \le G_{\rm max}$. It follows that
$P_{X^{\perp}} V \trialWF$ belongs to the finite-dimensional space
\begin{align}
  \label{eq:def_Y}
  Y = {\rm Span} \left\{e_{\bG}, \bG \in \mathcal R^{*}, \frac 1 2 |\bk+\bG|^{2}
  \le \Ecutt\right\}
\end{align}
with $\Ecutt = \frac 1 2 \left( \sqrt{2 \Ecut} + G_{\rm max} \right)^{2}$.
We therefore extend $\trialWF$ to this new basis $Y$ by zero-padding,
and compute $V \trialWF$ in this new basis, resulting in an exact
computation of the residual.
The very quick decay of the %
residual with increasing $\Ecut$ is shown in Figure \ref{fig:cb_residual}
for the first eigenvalue at the $\Gamma$ point.

\subsection{Goedecker--Teter--Hutter (GTH) pseudopotentials}
With the GTH pseudopotentials, $H \trialWF$ extends on all wave
vectors in ${\mathcal R}^*$, and therefore we cannot compute $\|P_{X^{\perp}} V \trialWF\|^2$
explicitly. Rather, as before, we compute $\|P_{Y} V \trialWF\|^2$,
where the larger approximation subspace $Y$, still defined by
\eqref{eq:def_Y} is determined by a chosen $\Ecutt > \Ecut$.
We then bound
the remaining term
\[
  \norm{P_{Y^{\perp}} V \trialWF}^2 = \sum_{\bG \in Y^{\perp}} \left| \sum_{\bG' \in X} \langle e_{\bG} | V | e_{\bG'} \rangle \trialWF(\bG')\right|^{2}.
\]
We do this by exploiting the fact that the matrix elements
$\langle e_{\bG} | V | e_{\bG'} \rangle$ connecting small wave vectors
in $\bG' \in X$ to large wave vectors in $\bG \in Y^{\perp}$ are
small. We obtain an explicit bound using our knowledge of $V$ and the
decay properties of Gaussians, see Appendix A for details.

\begin{figure}[h]
\centering
  \includegraphics[width=\columnwidth]{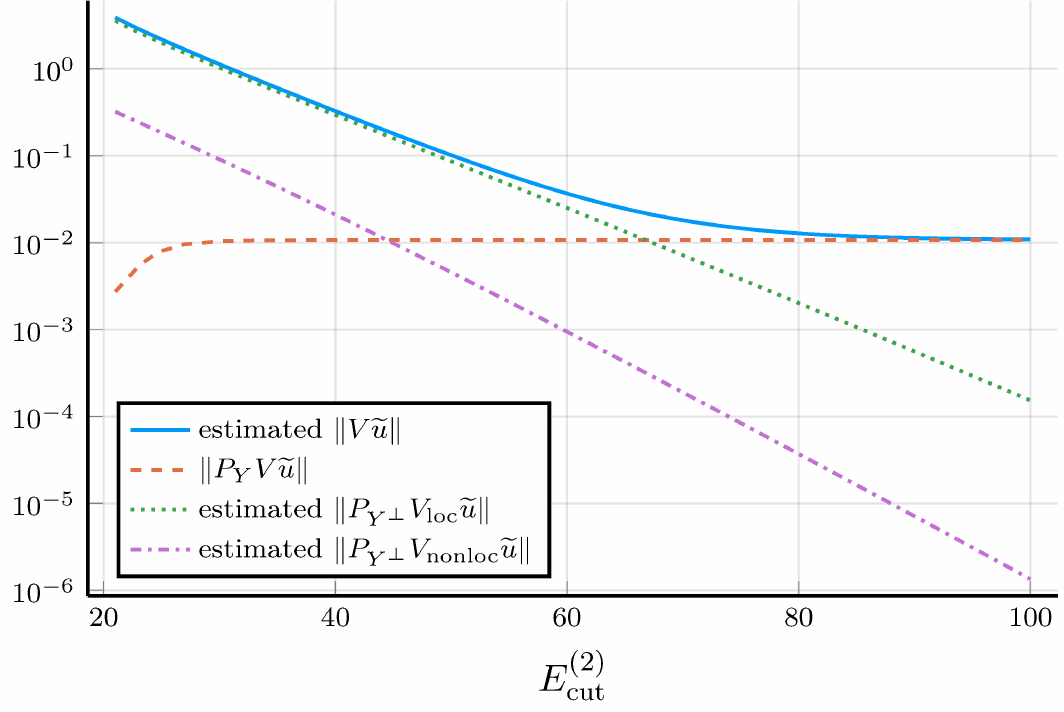}
  \caption{Bound on the residual $\|P_{X^{\perp}}V \widetilde u\|$
    and its three components for the first eigenvalue at the $\Gamma$
    point of the GTH model of Silicon as a function of $\Ecutt$. The initial
    $\widetilde u$ is obtained at $\Ecut=20$.
  }
  \label{fig:si_ecut2}
\end{figure}

Our rigorous bound of the residual is plotted Figure
\ref{fig:si_ecut2} along with its different components as a function
of $\Ecutt$, for an initial $\widetilde u$ obtained with $\Ecut = 20$.
As can be seen clearly,
the residual is essentially in $Y$ as soon as $\Ecutt \ge 30$,
our estimate, however, is only accurate for larger values of $\Ecutt \approx 80$.
This is since our bounds
for $\norm{P_{Y^{\perp}} V_{\rm loc} \trialWF}$
are still rather crude (see Appendix A).
Improving these is work in progress.

\begin{figure}[h]
\centering
  \includegraphics[width=\columnwidth]{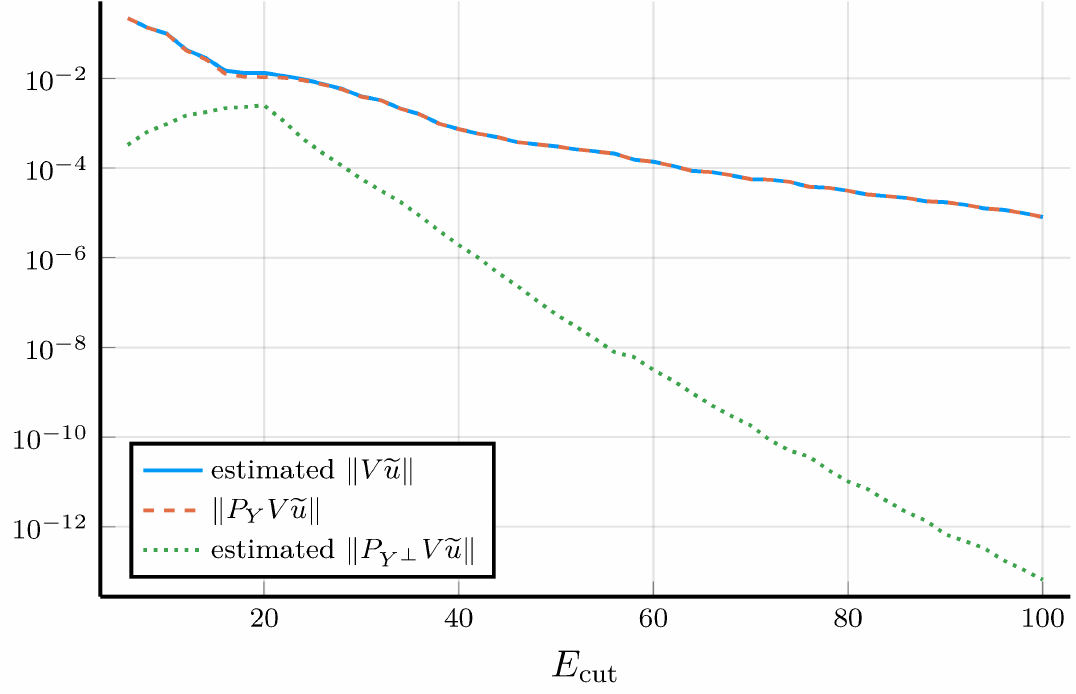}
  \caption{Computed and bounded terms of the residual norm
  for the first band at the $\Gamma$-point
  of the GTH model of silicon, using $\Ecutt = 4\Ecut$.
  The orange dashed and the solid blue curve are almost superimposed,
  indicating that the bounded $\norm{P_{Y^\perp} V \trialWF}$ term is negligible.
}
  \label{fig:si_residual}
\end{figure}

In theory, one could choose $\Ecutt$ dynamically based on the size of
the estimated residual. In the following, we use the simple heuristic
$\Ecutt = 4 \Ecut$, deduced from $\Ecut=20$ and Figure \ref{fig:si_ecut2}.
As can be seen in Figure~\ref{fig:si_residual},
$\Ecut=20$ represented the worst case for our heuristic.
With this choice of $\Ecutt$
the component on $Y^{\perp}$ of the residual is negligible
for all values of $\Ecut$ and our bound is nearly optimal.

\section{Estimating the gap}
The Kato-Temple bound \eqref{eq:KT}
requires a lower bound on the gap $\delta$ between the
(unknown) exact eigenvalue $\exactEV$ and the rest of the (unknown)
spectrum of $H$.
Let $N$ be the index of the target eigenvalue. Given an operator $H$
with eigenvalues $\exactEV_{1} \le \exactEV_{2} \le \dots$, how can we
obtain a lower bound on both $\exactEV_{N+1} - \exactEV_{N}$ in the
one direction and on $\exactEV_{N}-\exactEV_{N-1}$ in the other? In
the rest of this section we focus on a lower bound for
$\exactEV_{N+1} - \exactEV_{N}$, the other one being obtained
similarly.

Assume that we have computed the discretization $H_{XX}$ of the
Hamiltonian $H=\frac 12 (-\nabla + \bk)^2+V$ on a plane-wave basis set
$X$ with finite cutoff energy $\Ecut$, and let $\trialEV_{n,X}$ be its
eigenvalues with corresponding orthonormal eigenvectors
$\widetilde{u}_{n,X}$. Note that for the purposes of this section we
assume that all computations inside the basis set $X$ are exact: we
form explicitly the matrix representation of $H_{XX}$ and diagonalize
it using a dense eigensolver, and do not consider any arithmetic error.

\begin{figure}[h]
\centering
  \includegraphics[width=\columnwidth]{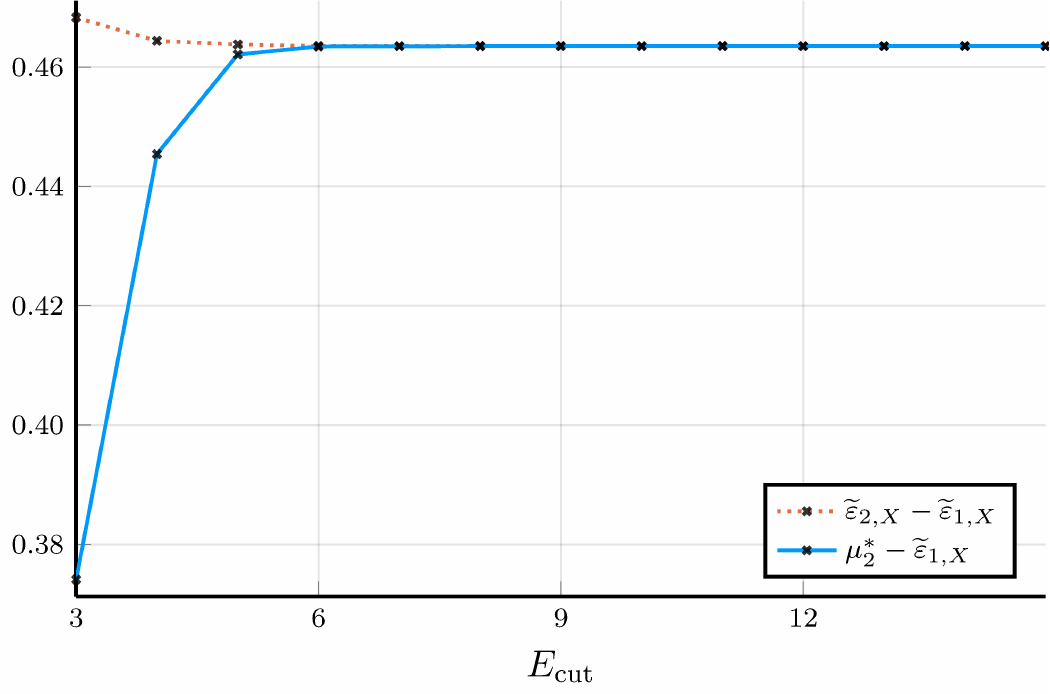}
  \caption{
	  Comparison of gap lower bounds for the Cohen-Bergstresser model of silicon.
	  The orange dotted is the na\"ive estimate $\trialEV_{2,X} - \trialEV_{1,X}$
	  between the computed approximate eigenvalues.
	  Shown in solid blue is $\mu^\ast_2 - \trialEV_{1,X}$ with $\mu$
	  a lower bound to $\trialEV_{2,X}$
	  obtained as described in the main text with $M=8$ eigenpairs.}
  \label{fig:cb_gap}
\end{figure}

From the variational principle, $\trialEV_{N,X} \ge \exactEV_{N}$. We
now need to obtain a rigorous lower bound on $\exactEV_{N+1}$, which
is is much more complex. Simply taking the difference
$\trialEV_{N+1,X} - \trialEV_{N,X}$ may lead to an
\emph{overestimation} of the gap,
see for example $\Ecut = 3$ in Figure \ref{fig:cb_gap}.
To compute a proper lower bound on
$\exactEV_{N+1}$ we express the operator $H$ on $X$ and its orthogonal
complement $X^{\perp}$ as
\begin{align*}
  H =
  \begin{pmatrix}
    H_{XX} & V_{X X^{\perp}}\\
    V_{X^{\perp} X} & H_{X^{\perp} X^{\perp}}
  \end{pmatrix}.
\end{align*}
In this we have used that the kinetic energy term is diagonal in a
plane-wave basis and does not appear in the off-diagonal blocks. We
then use the Haynsworth inertia additivity formula~\cite{Haynsworth}:
for any real threshold $\mu$ not in the spectrum of $H_{XX}$, the number of
eigenvalues of $H$ below $\mu$ is equal to the number
of negative eigenvalues of the block $H_{XX} - \mu$ plus the number of
negative eigenvalues of the Schur complement
\[
  S_{\mu} = (H_{X^{\perp} X^{\perp}} - \mu)
  - V_{X^{\perp} X} \left(H_{XX} - \mu\right)^{-1} V_{X X^{\perp}}.
\]

For pedagogical purposes we briefly sketch the proof of this
statement in the simplified case
where $H_{XX} - \mu$ is positive. Consider the quadratic form
\begin{align*}
  \langle  u, (H-\mu) u \rangle &= \langle  u_{X}, (H_{XX}-\mu) u_{X} \rangle + \langle  u_{X}, V_{XX^{\perp}} u_{X^{\perp}} \rangle \\
  &+ \langle  u_{X^{\perp}}, V_{X^{\perp}X} u_{X} \rangle + \langle  u_{X^{\perp}}, (H_{X^{\perp}X^{\perp}}-\mu) u_{X^{\perp}} \rangle
\end{align*}
where $u = (u_{X},u_{X^{\perp}})$. For a fixed $u_{X^{\perp}}$, this
is again a quadratic form in $u_{X}$, which can be explicitly minimized by
solving a linear system:
\begin{align*}
  \argmin_{u_{X} \in X}\, \langle  u, (H-\mu) u \rangle = -(H_{XX}-\mu)^{-1} V_{XX^{\perp}} u_{X^{\perp}}
\end{align*}
and therefore
  \begin{align*}
  \min_{u_{X} \in X} \langle  u, (H-\mu) u \rangle = \langle u_{X^{\perp}}, S_{\mu} u_{X^{\perp}}\rangle
\end{align*}
It follows that $S_{\mu}$ is positive if and only if $H-\mu$ is. The
extension to arbitrary number of negative eigenvalues proceeds
similarly.

From this, it follows that if we manage to prove that $S_{\mu}$ is
positive for some $\mu \in (\trialEV_{N,X},\trialEV_{N+1,X})$, we
obtain that $\exactEV_{N+1} \ge \mu$. Assume that we have computed
eigenvectors $\widetilde{u}_{1,X},\dots, \widetilde{u}_{M,X}$ up until
index $M \ge N+1$. Then the Schur complement $S_{\mu}$ can be bounded from below
by expanding $H_{XX}$ on its eigenvector basis $\trialWF_{n,X}$
\begin{equation}
  \begin{aligned}
    S_{\mu} &\ge \Ecut - \|V_{X^{\perp} X^{\perp}}\|_{\rm op} - \mu \\
    &\hspace{20pt}\underbrace{
      - \opnorm{
          \sum_{n=N+1}^{M-1}\,
          \frac{(V_{X^{\perp} X}\trialWF_{n,X}) (V_{X^{\perp} X}\trialWF_{n,X})^\dagger}{\trialEV_{n,X} - \mu}}}_{= B_\mu}
    - \frac{\opnorm{V_{X X^{\perp}}}^2}{\trialEV_{M,X} - \mu},
  \end{aligned}
  \label{eqn:SchurBound}
\end{equation}
where $\opnorm{\,\cdot\,}$ is the operator norm on the space
of bounded linear operators on $\mathcal H$.
Computing the term $B_\mu$
requires knowing the potential $V$ on all of the complement $X^\perp$,
which is not feasible for GTH pseudopotentials.
Similarly as with the residuals, we assume that we are able to compute $V$ on a
superset $Y \supset X$ and additionally able to bound it on $Y^\perp$.
With this in place we split $B_\mu$
into contributions inside and outside of $Y$:
\newcommand{\LL}{\widetilde{\Lambda}}
\newcommand{\UU}{\widetilde{U}}
\newcommand{\XcY}{X^{\perp}\cap Y}
\begin{equation}
	\begin{aligned}
	B_\mu \geq &-\opnorm{(V_{\XcY,X}\, \UU)\, (\LL - \mu)^{-1}\, (V_{\XcY,X}\, \UU)^{\dagger}} \\
		&-2 \opnorm{(V_{\XcY,X}\, \UU)\, (\LL - \mu)^{-1}}  \opnorm{V_{X,Y^\perp}}
		- \frac{\opnorm{V_{X,Y^\perp}}^2}{\trialEV_{N,X} - \mu},
	\end{aligned}
	\label{eqn:bBound}
\end{equation}
where $\LL$ is the diagonal matrix of eigenvalues
$\trialEV_{N+1,X}, \ldots, \trialEV_{M-1,X}$
and $\UU$ the orthogonal matrix of corresponding eigenvectors
(column-wise).

To compute the first term we use the fact that $V_{\XcY,X} \UU$ is a
computable, long-and-thin matrix (more rows than columns). We can
employ the $QR$ decomposition to factorize it into the product of an
orthogonal matrix $Q$ and a (small) triangular matrix $R$. The first
term in the right-hand side of \eqref{eqn:bBound} can then be explicitly computed as
the largest eigenvalue of the (small) Hermitian matrix
${R \, (\LL - \mu)^{-1}\, R^\dagger}$. The second term can be
treated similarly. Details on our bounds on the operator norms
$\opnorm{V_{X^{\perp} X^{\perp}}}$, $\opnorm{V_{X X^{\perp}}}$ and
$\opnorm{V_{X,Y^\perp}}$ for the Cohen-Bergstresser and the GTH
pseudopotential models are given in Appendix A.

For a fixed $\mu \in (\trialEV_{N,X}, \trialEV_{N+1,X})$, $S_{\mu}$ is
positive for all $\Ecut$ large enough.
For a given $\Ecut$ and $M$,
we find the best lower bound to $\trialEV_{N+1,X}$,
denoted by $\mu_{N+1}^\ast$,
as the maximum $\mu$, for which we can ensure that $S_{\mu}$ is positive.
This is done using a bisection algorithm on our bound of $S_{\mu}$.

\begin{figure}[h]
\centering
  \includegraphics[width=\columnwidth]{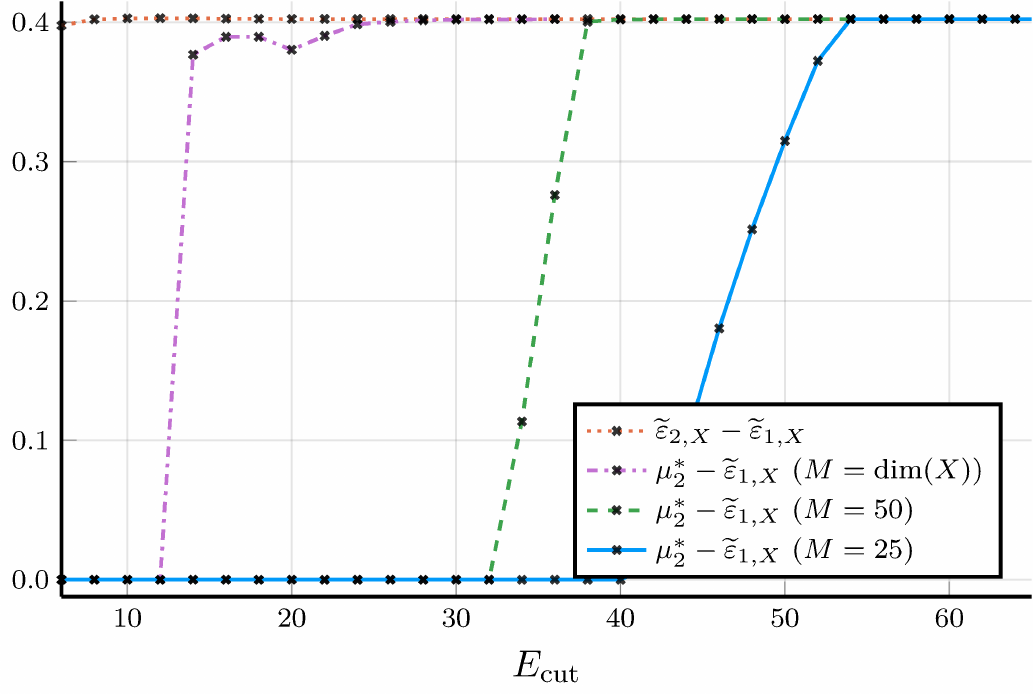}
  \caption{
	  Gap lower bounds of the GTH pseudopotential model of silicon.
	  Similar to Figure \ref{fig:cb_gap},
	  dotted orange indicates the na\"ive eigenvalue difference,
	  dashed green the lower bound $\mu^\ast_2 - \trialEV_{1,X}$
	  obtained using $M=50$ eigenpairs
	  and solid blue using $M=25$ eigenpairs.
	  The purple dashed line refers to the case where $H_{XX}$
	  is fully diagonalized, i.e. $M = \text{dim}(X)$.
	}
  \label{fig:si_gap}
\end{figure}

Figures \ref{fig:cb_gap} and \ref{fig:si_gap}
show gaps obtained using this lower bound on $\exactEV_{N+1}$
for different values of $\Ecut$ and $M$
for the first eigenvalue at the $\Gamma$ point
of the two pseudopotential models we consider. For the
simple Cohen-Bergstresser model and using $M=8$ our lower bound for the gap is
accurate already at $\Ecut = 10$. For the GTH pseudopotentials
one needs to use larger values of $M$ and $\Ecut$.

\section{Estimating the arithmetic error}
To rigorously bound the arithmetic error, we use interval arithmetic, as
specified in the IEEE standard 1788-2015~\cite{IEEEstandard}.
The main idea of interval arithmetic is to use
not one but two floating-point numbers to represent a given quantity,
forming an interval which contains the exact number. Every operation
is then performed on the limits of the interval utilizing rounding
modes of CPUs to ensure that the upper limit is only rounded upwards
and the lower limit only rounded downwards if rounding is needed to
represent the outcome of the operation. This ensures that the
\emph{exact} answer always lies between the limits of the final
interval. This simplistic approach generally overestimates
floating-point error, since it considers all floating-point operations
independent from each other. The bound obtained in this way is
thus too large making an application of interval arithmetic to,
e.g.~a complete iterative diagonalization algorithm, impractical.
Fortunately if we obtain an approximate eigenpair $(\trialEV, \trialWF)$ inside a
discretization basis $X$ using ordinary floating-point arithmetic, we
only need to re-evaluate the residual norm
$\norm{P_X (H\trialWF - \trialEV\trialWF)}$ in interval arithmetic to
obtain a rigorous bound. The upper bound of the resulting interval
gives access to the sum of \emph{both} floating-point error and
algorithmic error due to the eigensolver.

\begin{figure}[h]
\centering
  \includegraphics[width=\columnwidth]{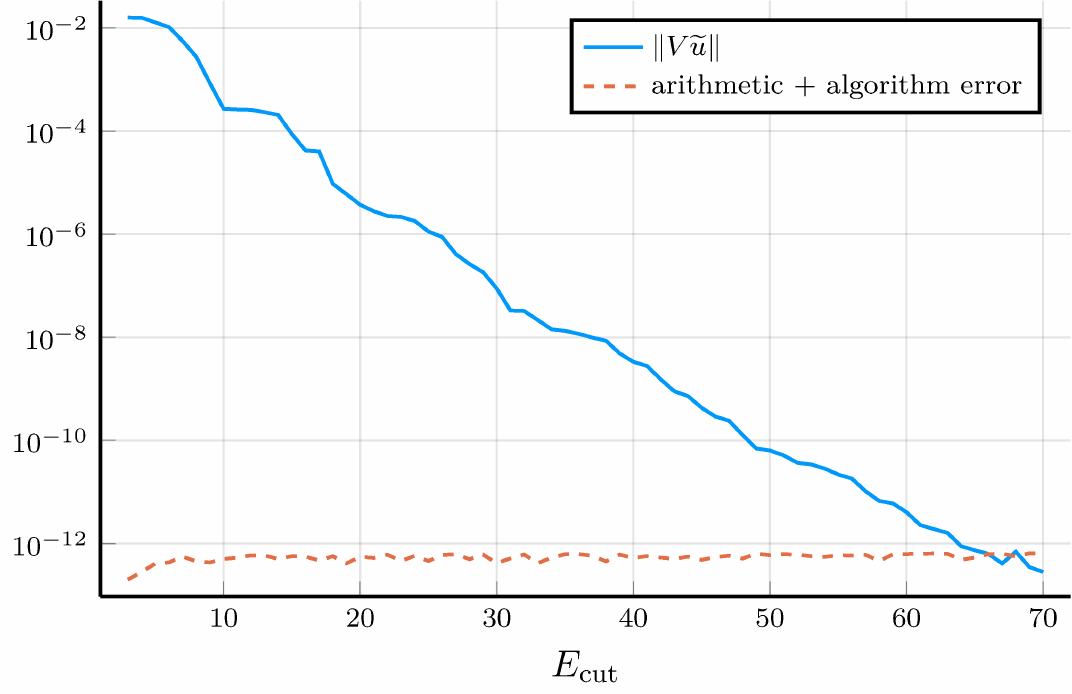}
  \caption{
	  Total residual and sum of algorithm and floating-point errors
	  at double-precision arithmetic
	  for the first band at the $\Gamma$ point of
	  the Cohen-Bergstresser model of silicon.
}
\label{fig:cb_arithmetic}
\end{figure}

In \julia IEEE interval arithmetic is available in the
\texttt{IntervalArithmetic.jl}~\cite{IntervalArithmetic} package as a
custom floating-point type. This type can be directly used with any
native \julia code, including
\texttt{GenericLinearAlgebra.jl}~\cite{GenericLinearAlgebra} and
\texttt{FourierTransforms.jl}~\cite{FourierTransforms}, which provide
interval-arithmetic equivalents to classical linear algebra and FFT algorithms.
These packages allowed us to use the routines of \dftk to also
bound the arithmetic error for all our double-precision
calculations presented in this paper. Example values are shown in
Figure \ref{fig:cb_arithmetic} indicating that the obtained arithmetic
error is several orders of magnitude smaller than the discretization
error until around $\Ecut = 65$. Increasing the accuracy of the
obtained solution to the Cohen-Bergstresser model beyond this point would not
only require to increase the basis, but also to switch to a more
accurate floating-point arithmetic beyond IEEE double precision. In a
type-generic \julia code like \dftk this is, however, completely
seamless and has been used to obtain Figure \ref{fig:cb_gamma},
demonstrating errors below double precision.

\section{Tracking errors in band computations}
\begin{figure}[h]
\centering
  \includegraphics[width=\columnwidth]{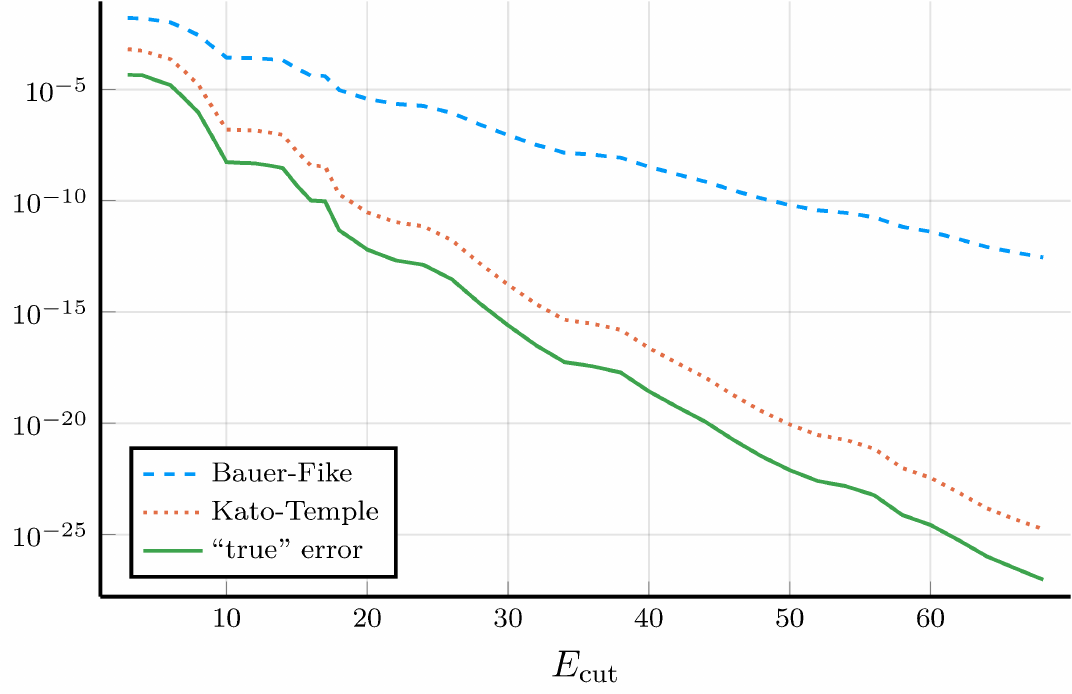}
  \caption{
	Overview of discretization error and error bounds at the $\Gamma$ point
	for the first band in the Cohen-Bergstresser model of silicon.
	The \texttt{DoubleFloats.jl}~\cite{DoubleFloats} package
	was used to go beyond the
	accuracy limitations of IEEE double precision.
	Gap lower bounds for the Kato-Temple bound
	have been obtained with $M=8$ eigenpairs.
	The ``true'' error was computed by taking the
	absolute difference to the eigenvalue
	obtained at $\Ecut = 70$.
  }
  \label{fig:cb_gamma}
\end{figure}
\begin{figure}[h]
\centering
  \includegraphics[width=\columnwidth]{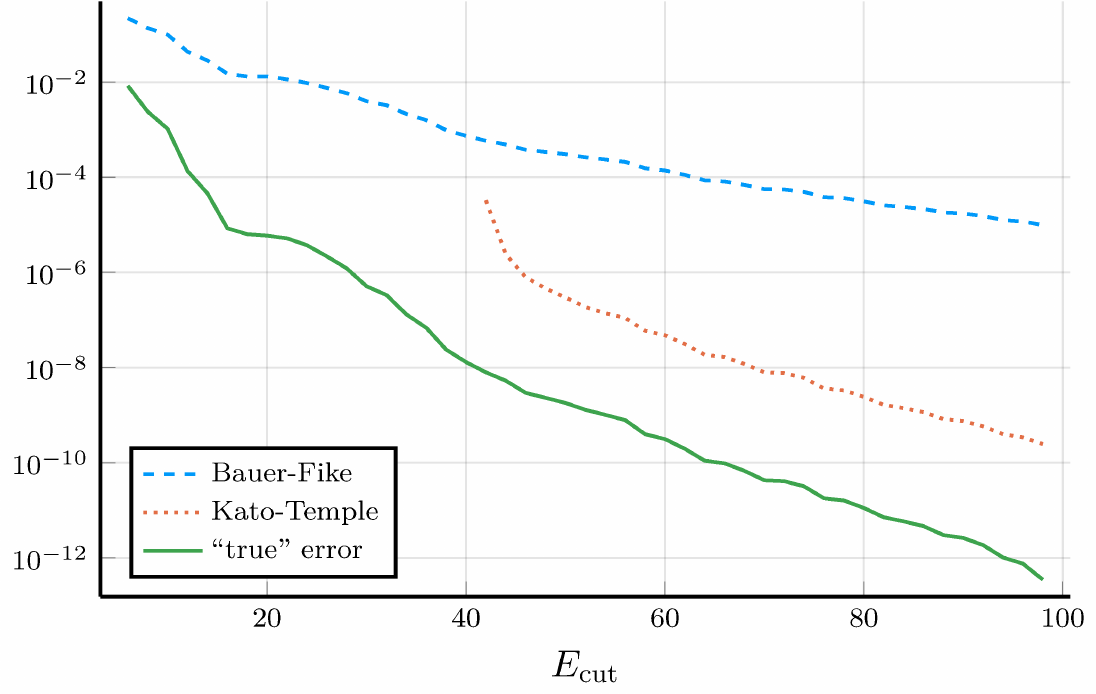}
  \caption{
	Overview of discretization error and error bounds at the $\Gamma$ point
	for the first band of the GTH model of silicon.
	Gap lower bounds for the Kato-Temple bound
	have been obtained with $M=25$ eigenpairs.
	The ``true'' error was computed by taking the
	absolute difference to the eigenvalue
	obtained at $\Ecut = 100$.
  }
  \label{fig:si_gamma}
\end{figure}
\begin{figure}[h]
\centering
  \includegraphics[width=\columnwidth]{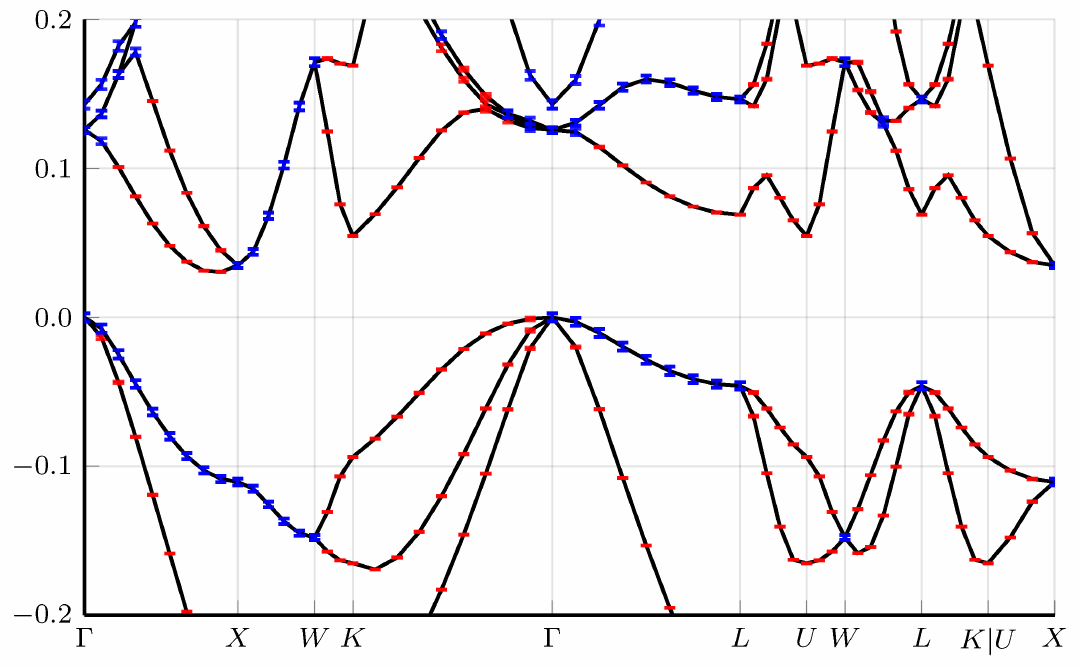}
  \caption{
	Band structure of silicon using the Cohen-Bergstresser model at $\Ecut=10$.
	Error bars indicate the total numerical error,
	i.e.~arithmetic error, algorithm error and discretization error.
	Red (resp. blue) error bars indicate that the Kato-Temple (resp. Bauer-Fike) bound gave the lowest
	discretization error and was used.
	Gap lower bounds have been obtained with $M=8$ eigenpairs.
	Band energies are given in Hartrees and relative
	to the $4$th eigenvalue at the $\Gamma$ point.
  }
  \label{fig:cb_bands}
\end{figure}
\begin{figure}[h]
\centering
  \includegraphics[width=\columnwidth]{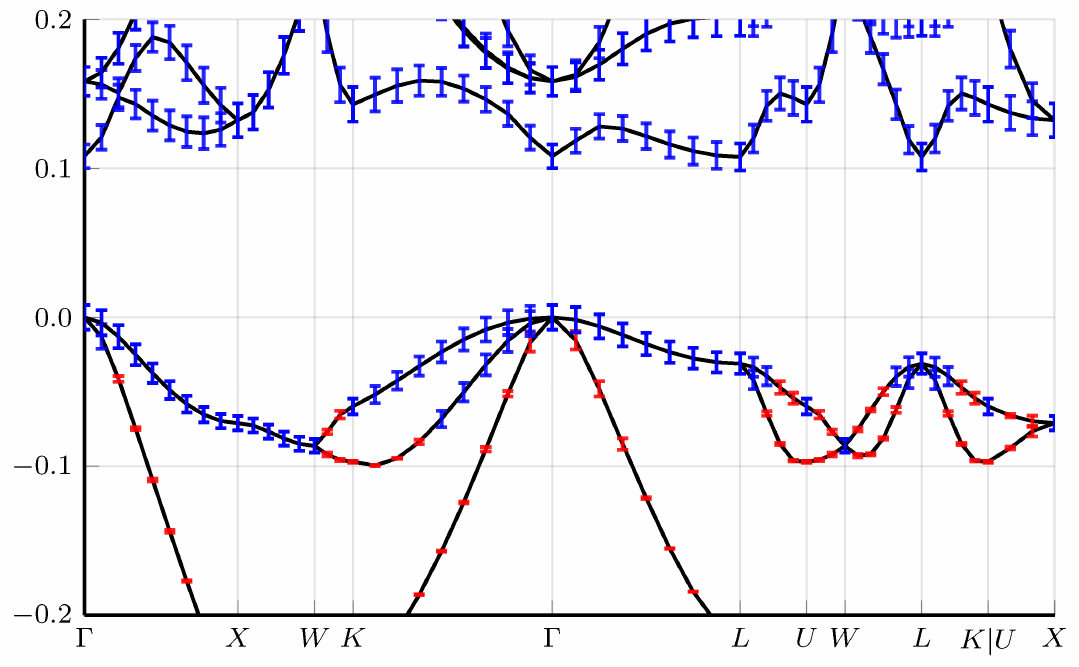}
  \caption{
	Band structure with error bars for the GTH model of silicon
	at $\Ecut = 42$.
	The same conventions as in Figure \ref{fig:cb_bands} are used,
	except that gap lower bounds have been obtained using $M = 35$ eigenpairs.
  }
  \label{fig:si_bands}
\end{figure}

With the discussed methodologies we are able to
\begin{itemize}
\item compute the norm of the in-space residual norm $\|P_X \res\|$ (leading to the algorithmic error);
\item compute a guaranteed upper bound of the out-of-space residual norm $\|P_{X^\perp} \res\|$ (leading to the discretization error);
\item compute a guaranteed lower bound of the spectral gap $\delta$ involved in the Kato-Temple inequality~\eqref{eq:KT};
\item compute {\it a posteriori} errors bounds on the quantities of interest (crystalline orbital energies) using either Bauer-Fike or Kato-Temple inequalities;
\item estimate the impact of floating-point errors via interval arithmetic (arithmetic error).
\end{itemize}
A summary of the overall numerical error on the quantities of interest
bound by Bauer-Fike and Kato-Temple versus the ``true'' error
estimated using the largest basis employed ($\Ecut = 50$ and
$\Ecut = 100$ respectively) is given in Figures \ref{fig:cb_gamma} and
\ref{fig:si_gamma} for the Cohen-Bergstresser and GTH models of
silicon. Using our methods we were able to compute band structures for
both models with fully guaranteed bounds on the numerical error, see
Figures \ref{fig:cb_bands} and \ref{fig:si_bands}, respectively.
Notice that neither Kato-Temple nor Bauer-Fike universally provide the
best bound on the error, with the Kato-Temple bound being inapplicable
for degenerate eigenvalues in particular. In our plots we alternate
between both bounds, only showing the one providing the smallest
error. The design and use of error bounds robust to degenerate
eigenvalues is left to future work.


\section{Conclusions and outlook}
We have discussed the computation of fully guaranteed bounds to the
error in the numerical solution of non-self-consistent Kohn-Sham
equations. While many other sources of error exist, our treatment
focused exclusively on the discretization error, due to finite basis
sets, algorithm error, due to non-zero convergence thresholds, and
arithmetic error, due to finite-precision floating-point arithmetic.
As quantities of interest we considered the band energies around the
Fermi level of the Cohen-Bergstresser and GTH pseudopotential models
and demonstrated the feasibility of our approach by providing, for
each model, band structure diagrams of silicon, in which the total
numerical error was annotated in the form of error bars. In this paper
we have focused on error bounds on eigenvalues for simplicity, but
that is not a major limitation of our approach; error bounds on
eigenvectors can be obtained in a similar fashion to the Kato-Temple
bound (see \cite{saad}, Theorem 3.9). From there it is possible to
obtain bounds on other quantities of interest, for instance the
density or the forces.

Our presented approach relies on two key ingredients. First, a bound
on the residual, including both its in-space component (the one driven
to zero by the iterative solvers) and its out-of-space component. For
the out-of-space component, we relied in this work on the properties
of plane-wave basis and of the analytical pseudopotential; however,
this can in principle be applied to any basis set for which exact
analytical computations are possible, such as Gaussian basis sets. The
second ingredient is to relate the residual to the actual error, which
involved for our particular quantities of interest an eigenvalue gap.
This is much more complex to perform in full rigor, as it requires a
control of the out-of-basis part of the full operator itself.
Achieving a similar analysis to ours in Gaussian basis sets (for
instance) is a major research challenge.

A limitation of our study is that we only considered
non-self-consistent models. Taking into account self-consistency in
error bounds one runs into two separate issues. First, the non-convexity
of the model due to the exchange-correlation term can introduce
multiple energy local minima, and it is impossible to certify that the
true ground state has been found. One then has to settle for a less
ambitious notion of error control: showing that there exists a
solution of the equation close to the numerically obtained approximate
solution. Second, rigorously proving error bounds for nonlinear
problems involves mathematically more sophisticated techniques to
achieve a sufficient control on the nonlinear terms. This has been
performed for the Gross-Pitaevskii equation, an equation similar in
form but simpler than Kohn-Sham DFT \cite{dusson2017posteriori}.
Extending this to Kohn-Sham DFT is work in progress.

Finally, we hope to use our bounds to enable a fully black-box
modeling, where accuracy parameters such as the kinetic energy cutoff,
the convergence thresholds of the iterative solver or employed
floating-point precision are chosen automatically by the code. The
hope is to be able to dynamically adjust such accuracy parameters
during a simulation to do as little work as necessary to reach the
accuracy desired by the user. For such purposes we expect the
presented Kato-Temple-type bounds to be not tight enough, so that
better bounds have to be constructed and implemented. Preliminary work
in that direction can be found in~\cite{CDMSVclusters}.

\section*{Conflicts of interest}
There are no conflicts of interest to declare.

\section*{Acknowledgements}
This project has received funding from
ISCD (Sorbonne Universit\'e) and from the European Research Council (ERC) under
the European Union's Horizon 2020 research and innovation program (grant
agreement No 810367).

\section*{Appendix A: Bounds on potentials}
In this section we fix the two plane-wave spaces $X$ and $Y$ defined
by energy cutoffs $\Ecut$ and $\Ecutt > \Ecut$.
Our error bounds for the GTH pseudopotentials
require explicit bounds on the quantities $\|P_{Y^{\perp}} V \trialWF\|$
for a given $\trialWF \in X$ (to bound the residuals), and on the operators
$P_{X^{\perp}} V P_{X}$,
$P_{Y^{\perp}} V P_{X}$,
and $P_{X^{\perp}} V P_{X^{\perp}}$ (to obtain
bounds on the gap).

We will bound these quantities using our explicit knowledge of the tails of the functions
involved: given a non-negative continuous function $f:\R_+ \to \R_+$, we define
\begin{align*}
  S_{\cR^*}(f, q) = \sum_{\bG \in \mathcal R^{*}, |\bG| \ge q} f(|\bG|).
\end{align*}
Bounds on $S_{\cR^*}(f, q)$ are obtained in Appendix B.

Note that there is a large freedom in designing bounds, and many
possible improvements. The ones we present result from a compromise
between accuracy and simplicity.

\subsection*{Local potential}
In the following we assume a single atom centered at the origin; the
case of multiple atoms is obtained simply by adding a contribution for
each atom. The function $\widehat v_{\rm loc}(\Delta \bG)$ is then radial.%

To bound $P_{Y} V_{\rm loc} \trialWF$ for a given $\trialWF$, we use the fact
that $\trialWF$ has small components on high wavevectors:
\begin{align*}
  &|\Omega|^{2} \|P_{Y^{\perp}} V_{\rm loc} \trialWF\|^{2}\\
  &= \sum_{\bG \in {Y^\perp}} \left| \sum_{\bG' \in X} \widehat v_{\rm loc}(\bG-\bG') \trialWF(\bG')\right|^{2}\\
  &\le {\rm dim}(X) \sum_{\bG \in {Y^\perp}} \sum_{\bG' \in X} \left| \widehat v_{\rm loc}(\bG-\bG') \trialWF(\bG')\right|^{2}\\
  &= {\rm dim}(X) \sum_{\bG' \in X} |\trialWF(\bG')|^{2} \sum_{\bG \in {Y^\perp}} \left| \widehat v_{\rm loc}(\bG-\bG')\right|^{2}\\
  &\le {\rm dim}(X) \sum_{\bG' \in X} |\trialWF(\bG')|^{2} \sum_{\substack{\Delta \bG \in \mathcal R^{*}\\ |\Delta \bG| > \sqrt{2 \Ecutt} - |\bG'|}}  |\widehat v_{\rm loc}(|\Delta \bG|)|^{2}\\
  &= {\rm dim}(X) \sum_{\bG' \in X} |\trialWF(\bG')|^{2} \, S_{\cR^*}\left(|\widehat v_{\rm loc}|^{2}, \sqrt{2 \Ecutt} - |\bG'|\right)
\end{align*}

We bound the operators
$P_{X^{\perp}} V_{\rm loc} P_{X}$, $P_{Y^{\perp}} V_{\rm loc} P_{X}$
and $P_{X^{\perp}} V_{\rm loc} P_{X^{\perp}}$ in operator norm simply
by the operator norm of $V_{\rm loc}$. We use the fact that $V_{\rm
  loc}$ is a multiplication operator in real space by the periodic function
\begin{align*}
  v_{\rm loc}(\br) = \frac 1 {|\Omega|} \sum_{\bG \in \mathcal R^{*}} \widehat{v}_{\rm loc}(\bG) e^{i \bG \cdot \br}
\end{align*}
and therefore
\begin{align*}
  \|V_{\rm loc}\|_{\rm op} &\le \frac 1 \Omega\left( \left\|\sum_{\bG \in Y} \widehat{v}_{\rm loc}(\bG) e^{i \bG \cdot \br}\right\|_{\infty} +  \sum_{\bG \in Y^{\perp}} |\widehat{v}_{\rm loc}| \right)\\
  &\le \frac 1 \Omega\left( \left\|\sum_{\bG \in Y} \widehat{v}_{\rm loc}(\bG) e^{i \bG \cdot \br}\right\|_{\infty} +  S_{\mathcal R^{*}}(|\widehat v_{\rm loc}|, \sqrt{2 \Ecut}) \right),
\end{align*}
where $\|v\|_{\infty} = \sup_{\br \in \R^{3}} |v(\br)|$. To bound the
\newcommand{\Gr}{\mathfrak{G}}
first term, we use the fact that for a regular grid $\Gr$ of the unit cell, we have
\begin{align*}
  \left\|\sum_{\bG \in Y} \widehat{v}_{\rm loc}(\bG) e^{i \bG \cdot \br}\right\|_{\infty} \le \max_{\br \in \Gr} \left|\sum_{\bG \in Y} \widehat{v}_{\rm loc}(\bG) e^{i \bG \cdot \br}\right| + \delta \sum_{\bG \in Y} |\bG| |\widehat{v}_{\rm loc}(\bG)|
\end{align*}
where $\delta$ is the diameter of the grid. We compute the first term
using a fast Fourier transform and the second explicitly.

\subsection*{Nonlocal potential}
The nonlocal potential is a sum of separable terms~\cite{GTH,HGH}, over
atoms, angular momentum $l$, magnetic quantum number $m$ and projector
channels $i,j$. We focus on just one of the terms in~\eqref{eq:Vnl}, of the form
\begin{align*}
  \langle  e_{\bG}|v_{\nl}|e_{\bG'} \rangle= p^{(1)}(\bk+\bG) \, \overline{p^{(2)}(\bk+\bG')},
\end{align*}
where both $p^{(i)}$
are of the form $p^{(i)}(\bq) = Y_{lm}(\bq/|\bq|) R^{(i)}(|\bq|)$ ($i = 1,2,\ldots$).
A bound on the total
potential can be obtained naively by the triangular inequality. A
better bound can be obtained by exploiting orthogonality between the
different quantum numbers $(l,m)$ and using Pythagoras theorem
as well as using Uns\"old's theorem to simplify the sums over $m$.
We do not detail these technicalities here.

For a given $\trialWF \in X$, let $c_{\widetilde u} = \sum_{\bG \in X}
\overline{p^{(2)}(\bk+\bG)}\widetilde u(\bG)$. Then
\begin{align*}
\|P_{Y^{\perp}} v_{\nl} \trialWF\|^{2}&= |c_{\widetilde u}|^{2} \sum_{\bG \in {Y^\perp}} |p^{(1)}(\bk+\bG)|^{2}\\
  &\le |c_{\widetilde u}|^{2} \|Y_{lm}\|_{\infty}^2  \sum_{\bG \in {Y^\perp}} |R^{(1)}(|\bG| - |\bk|)|^{2}\\
  &= |c_{\widetilde u}|^{2}\|Y_{lm}\|_{\infty}^2  \, S_{\cR^*}\left(|R^{(1)}(\cdot-|\bk|)|^2, \sqrt{2 \Ecutt}\right),
\end{align*}
where
$\|Y_{lm}\|_{\infty} = \sup_{\bq \in \mathbb R^{3}, |\bq| = 1}
|Y_{lm}(\bq)|$ and we have assumed that $|R^{(i)}|$ is non-increasing on
$[\Ecutt - |\bk|,+\infty)$. We compute $c_{\widetilde u}$ explicitly and
bound the remainder $S$ as before. For the operator norm of $P_{X^{\perp}} V_{\nl} P_{X}$ and
$P_{Y^{\perp}} V_{\nl} P_{X}$, we use the above computation together
with the fact that, for any normalized $\widetilde u \in X,$
\begin{align*}
  |c_{\widetilde u}|^{2} \le \sum_{\bG \in X} |p^{(2)}(\bk+\bG)|^{2}.
\end{align*}
Finally,
\begin{align*}
    \|P_{X^{\perp}} V_{\nl} P_{X^{\perp}}\|^{2}_{\rm op}
  \le \|Y_{lm}\|_{\infty}^4\ \ \  &{S_{\cR^*}(|R^{(1)}(\cdot-|\bk|)|^{2}, \sqrt{2 \Ecut})}\\
  & {S_{\cR^*}(|R^{(2)}(\cdot-|\bk|)|^{2}, \sqrt{2 \Ecut})}.
\end{align*}

\section*{Appendix B: Bounds on tail sums}
\label{appendix:gaussian_bounds}
Our bounds involve quantities of the form
\begin{align*}
  S_{\cR^*}(f,q) := \sum_{\bG \in \mathcal R^{*}, |\bG| \ge q} f(|\bG|).
\end{align*}
where $f:\R_+ \to \R_+$ is a known smooth non-negative function non-increasing  on the internal $[q_{\rm min},+\infty)$ for some $q_{\rm min} \ge 0$, and more
specifically a Gaussian times a polynomial or rational function. We
seek to bound $S_{\cR^*}(f,q)$ by an explicitly computable quantity for $q$ large enough. For a one dimensional
lattice $\cR^*=b\Z$, $b>0$, this can easily be done using a
sum-integral comparison. We have for all $q \ge q_{\rm min} + b$,
$f(q)\le b^{-1} \int_{q-b}^q f(q') \, dq'$, so that
\begin{align*}
S_{b\Z}(f,q)  \le 2 b^{-1} \int_{q-b}^{\infty} f(q') \, dq',
\end{align*}
where the right-hand side can be computed explicitly using integrals
of Gaussian functions.

\begin{figure}[h]
\centering
\includegraphics[width=0.4\textwidth]{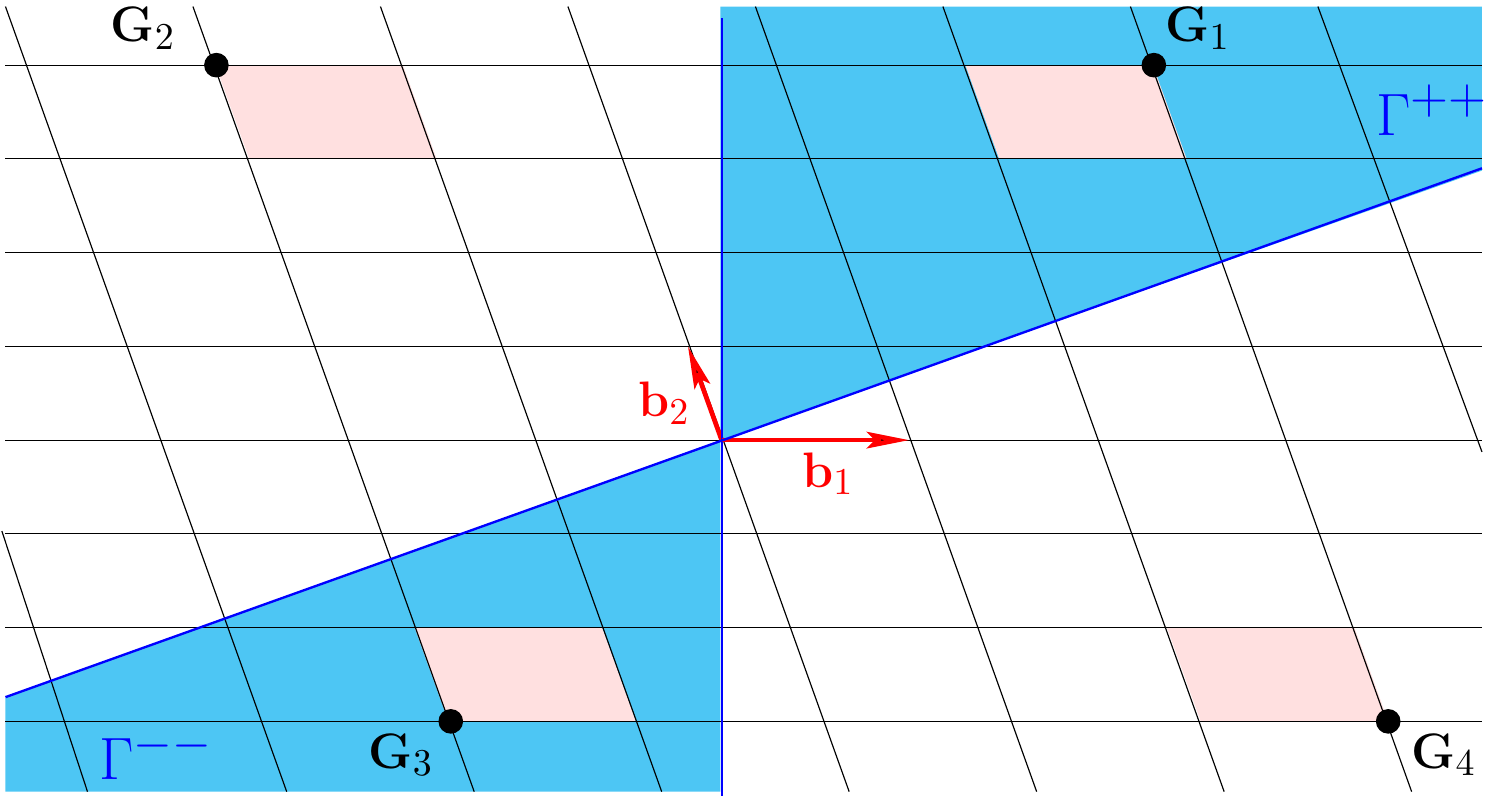}
\caption{Partitioning of the space used to decide in which direction
  to extend the unit cell $C_{\bG}$ (red).}
  \label{fig:2D_lattice}
\end{figure}

For the multidimensional case we use a similar idea: we bound the
value of $f(|\bG|)$ by its mean over a unit cell $C_\bG$ for which
$|\bG|$ is the vertex of $C_\bG$ furthest away from the origin (see
Figure \ref{fig:2D_lattice}). A
technical difficulty lays in the fact that this $C_\bG$ is not always
uniquely defined and that the $C_\bG$'s do not form a proper tiling.
For instance, for the two-dimensional square lattice, we have
\begin{align}
S_{b\Z^2}(f,q) &=  4 \sum_{\bG \in b (\N^* \times \N^*), |\bG| \ge q} f(|\bG|) + 2 \sum_{\bG \in b (\Z \times \{0\}), |\bG| \ge q} f(|\bG|) \nonumber \\
& \le 4 \sum_{\bG \in b (\N^* \times \N^*), |\bG| \ge q} \int_{C_{\bG}} f(\bG') d\bG' + 2 S_{b\Z}(f,q).\nonumber\\
& \le   2 \pi b^{-2} \int_{q - \sqrt 2 b}^{+\infty} q' f(q') \, dq' + 2 S_{b\Z}(f,q).\nonumber
\label{eq:bound_2D_square}
\end{align}
with a similar bound for the cubic lattice. For a non-cubic lattice $\cR^\ast=\bb_1 \Z + \bb_2 \Z + \bb_3 \Z$,
a partitioning in octants has to be done instead based on the
signs of the quantities $\bG \cdot \bb_{i}$, in the spirit of what can be seen on
Figure~\ref{fig:2D_lattice} for a two-dimensional non-square case.
Unit cells then overlap in the vicinity of the three planes
$\bb_{1}^{\perp}, \bb_{2}^{\perp} $ and $\bb_{3}^{\perp}$. A relatively straightforward
but tedious bound on these extra overlaps yields the following:
\begin{align}
S_{\cR^*}(f,q) &\le  \frac{4\pi}{|C|} \int_{q-\delta}^{+\infty} {q'}^2 f(q') \, dq' + 2\pi \left( \sum_{j=1}^3  \frac{n_{\cR^\ast,j}}{|C_j|}\right)  \int_{q-\delta}^{+\infty} q' f(q') \, dq' \nonumber \\
 & \quad  +
 2 \left( \sum_{j=1}^3 \frac{ n_{\cR^\ast,j} }{|\widetilde C_j|} \right) \int_{q-\delta}^{+\infty} f(q') \, dq',
\end{align}
with $\delta$ the unit cell diameter and
\begin{align*}
|C|&=|(\bb_1 \times \bb_2)\cdot\bb_3| \\
|C_j|&= \left| \left( \bb_{j+1} -
    \frac{(\bb_j\cdot\bb_{j+1})}{|\bb_j|^2} \bb_j \right) \times
  \left( \bb_{j+2} - \frac{(\bb_j\cdot\bb_{j+2})}{|\bb_j|^2} \bb_j
  \right) \right|\\
 |\widetilde C_j|&=\left| \bb_j - \sum_{k \neq j} \frac{(\bb_k\cdot\bb_j)}{|\bb_k|^2} \bb_k \right|\\
n_{\cR^*,j} &= \prod_{k \neq j} \big\lceil{2+\frac{|\bb_j \cdot \bb_k|}{|\bb_j|^2}}\big\rceil.
\end{align*}
with the convention that $\bb_{4}=\bb_{1}, \bb_{5}=\bb_{2}$.
Details of this computation will be published in an upcoming paper.

The bound above is numerically observed to be very pessimistic because
it uses values of $f(q')$ for $q' < q$, which for a rapidly decaying
function are much larger than $f(q)$. In order to obtain a better
error, we instead apply this bound to the function
$\widetilde f(q') = \min(f(q),f(q'))$.

\bibliography{literature}

\end{document}